\documentclass[11pt]{article}
\usepackage{amssymb,amsmath,epsfig,graphics,color,cite}
\def\one{{{{\rm 1} \kern -.19em {\rm l}}}}
\def\C{{{{\rm {\mbox{\small l}}} \kern -.50em {\rm C}}}}
\def\R{{{{\rm l} \kern -.15em {\rm R}}}}
\def\N{{{{\rm l} \kern -.15em {\rm N}}}}
\def\E{{{{\rm l} \kern -.15em {\rm E}}}}
\def\P{{{{\rm l} \kern -.15em {\rm P}}}} 
\def\Z{{{{\rm Z} \kern -.35em {\rm Z}}}}
\def\1{{{{\rm 1} \kern -.35em {\rm 1}}}}

\begin{document}
\vspace*{0cm}
\begin{center}
{\setlength{\baselineskip}{1.0cm}{ {\Large{\bf 
THE GENERALIZED ZERO-MODE SUPERSYMMETRY SCHEME AND THE CONFLUENT ALGORITHM
 \\}} }}
\vspace*{1.0cm}
{\large{\sc{Alonso Contreras-Astorga}$^\dagger$} and {\sc{Axel Schulze-Halberg}$^\ddagger$}}
\indent \vspace{.3cm} 
\noindent \\	
$^\dagger$ Department of Mathematics and Actuarial Science, Indiana
University Northwest, 3400 Broadway, Gary IN 46408, USA, e-mail: 
aloncont@iun.edu \\[1ex]

$\ddagger$ Department of Mathematics and Actuarial Science and Department of Physics, Indiana
University Northwest, 3400 Broadway, Gary IN 46408, USA, e-mail:
axgeschu@iun.edu, xbataxel@gmail.com 
\end{center}

\vspace*{.5cm}
\begin{abstract}
\noindent
We show the relationship between the mathematical framework used in recent papers by H.C. Rosu, S.C. Mancas and P. Chen (2014)
\cite{rosu1,rosu2,rosu3} and the second-order confluent supersymmetric quantum mechanics. In addition, we point out several immediate generalizations of the approach 
taken in the latter references. Furthermore, it is shown how to apply the generalized scheme to the Dirac and to the Fokker-Planck equation.

\end{abstract} 

\section{Introduction}
In their recent work \cite{rosu1,rosu2,rosu3}, Rosu et al. present and discuss several interesting examples of 
quantum potentials and their associated zero-energy solutions that are constructed through a method inspired by the 
quantum-mechanical supersymmetry (SUSY) formalism \cite{mielnik}. In general, the latter formalism provides a method for the construction 
of solvable spectral problems that are governed by the Schr\"odinger equation. There is a vast amount of literature on 
the general topic, for a general overview the reader may refer to the reviews \cite{cooper,Fernandez05,fernandezreview}. 
The general SUSY formalism splits into two different algorithms, which are usually referred to as the 
standard algorithm and the confluent algorithm. The standard algorithm is a direct generalization of the Darboux--Crum transformation \cite{Darboux82,Crum55} whereas the confluent case can be seen as ingenious iterations of the mentioned transformation. Both algorithms serve the same purpose, but they are technically 
inequivalent and as such lead to different results. Despite being less 
known than its standard counterpart, the confluent SUSY algorithm has been subject to studies in a variety of 
articles. The first applications of this algorithm are presented in \cite{baye2,baye1,Stahlhofen,Rosu98,Mielnik00,Rosu00}, while for recent 
progress the reader may refer to \cite{dj03,fernandezconfluent,dj05,dj11,dj12,xbatconfluent,contreras14} and references therein.

The main purpose of this work is to show the close relationship between the mathematical method used in 
\cite{rosu1,rosu2,rosu3}, which for the sake of simplicity we will call ``Zero-mode SUSY scheme$"$, and the second-order confluent SUSY algorithm. 

In section \ref{Section SUSYCon} we will present the second-order confluent SUSY algorithm and its relationship with the zero-mode SUSY scheme, also we will comment on some important features that the zero-mode SUSY scheme could take from the confluent SUSY algorithm. The focus of section \ref{Section application} is to apply and extend the results of the previous section to the Schr\"odinger, Dirac and Fokker-Planck equations. Finally, in section \ref{Section remarks} our concluding remarks will be presented.

\section{Second-order confluent SUSY algorithm}\label{Section SUSYCon}

In the following we will develop the confluent, second-order SUSY algorithm in order to show its close relationship with the zero-mode SUSY scheme, also we will point some important properties of the confluent algorithm that can generalize the zero-mode scheme. To this end, let us start out with the Schr\"odinger equation
\begin{eqnarray}
\Psi''+(\epsilon-V_1)~\Psi &=& 0, \label{eq1}
\end{eqnarray}
where the prime stands for the derivative with respect to the position,  $\epsilon$ for the stationary energy and $V_1$ denotes a time independent potential. We will now 
apply a first-order SUSY-transformation to the solution $\Psi$ of (\ref{eq1}), given by
\begin{eqnarray}
\bar{\Psi} &=& -\frac{u'}{u}~\Psi+\Psi', \label{susy1}
\end{eqnarray}
where the transformation function $u$ is a solution of 
\begin{eqnarray}
u''+(\lambda-V_1)~u &=& 0. \label{aux1}
\end{eqnarray}
This is the same equation as (\ref{eq1}), but for a stationary energy $\lambda$ (called factorization energy) 
that is allowed to be different from $\epsilon$. Observe further that $\Psi$ and $u$ must be linearly independent in order to avoid 
vanishing of (\ref{susy1}). The function $\bar{\Psi}$ then solves the Schr\"odinger equation
\begin{eqnarray}
\bar{\Psi}''+(\epsilon-V_2)~\bar{\Psi} ~=~ 0, ~~~~~~\mbox{where}\quad V_2 ~=~ V_1-2~\frac{d^2}{dx^2}~ \log\left(u\right). \label{eq2}
\end{eqnarray}
In the next step we apply a first-order SUSY transformation to the new equation (\ref{eq2}), which requires another 
transformation function $\bar{u}$ to generate the function
\begin{eqnarray}
\hat{\Psi} &=& -\frac{\bar{u}'}{\bar{u}}~\bar{\Psi}+\bar{\Psi}'. \label{susy2}
\end{eqnarray}
Let us choose the factorization energy of $\bar{u}$ to be the same 
$\lambda$ as for $u$ in (\ref{aux1}), such that $\bar{u}$ must solve
\begin{eqnarray}
\bar{u}''+(\lambda-V_2)~\bar{u} &=& 0. \label{aux2}
\end{eqnarray}
Our iteration of the two transformations (\ref{susy1}), (\ref{susy2}) is referred to as confluent SUSY algorithm, 
because the factorization energy is maintained constant \cite{Mielnik00}. While in the present case we are performing a 
transformation of second-order (two iterations), the confluent SUSY algorithm can be applied for 
arbitrarily high order \cite{xbatconfluent}. Now, the function $\hat{\Psi}$ must be a solution of the following Schr\"odinger equation:
\begin{eqnarray}
\hat{\Psi}''+(\epsilon-V_3)~\hat{\Psi} ~=~ 0, ~~~~~~\mbox{where}\quad V_3 ~=~ V_2-2~\frac{d^2}{dx^2}~ \log\left(\bar{u}\right). 
\label{eq3}
\end{eqnarray}
In order to construct the potential $V_3$, we recall that a particular solution of equation (\ref{aux2}) 
is given by $\bar{u}_p=1/u$. Application of the 
reduction-of-order formula then yields the general solution of (\ref{aux2}) in the form
\begin{eqnarray}
\bar{u} &=& c_1~\bar{u}_p+c_2~\bar{u}_p~\int\limits^x \frac{1}{\bar{u}^2_p}~dt ~=~
\frac{c_1}{u}+\frac{c_2}{u}~\int\limits^x u^2~dt, \label{baru}
\end{eqnarray}
where $c_1$ and $c_2$ are arbitrary constants. Note that in order to obtain \eqref{baru} we use the fact that the differential equation \eqref{aux2} does not contain a term with first order derivative, in these cases the Wronskian of the two linearly independent solutions is constant, and without loss of generality we set this constant equal to one since it can be absorbed in the coefficient $c_2$. We can now use the representation of $V_2$
in (\ref{eq2}) and the explicit form (\ref{baru}) to determine the 
transformed potential $V_3$ as follows.
\begin{eqnarray}
V_3 &=& V_2-2~\frac{d^2}{dx^2}~ \log\left(\bar{u}\right) \nonumber \\[1ex]
&=& V_1 -2~\frac{d^2}{dx^2}~ \log\left(u\right)-2~\frac{d^2}{dx^2}~ \log\left(\bar{u}\right) \nonumber \\[1ex]
&=& V_1 -2~\frac{d^2}{dx^2}~ \log\left(\frac{c_1}{c_2}+\int\limits^x u^2~dt \right),\label{v3}
\end{eqnarray}
note that every ratio $c_1/c_2$ defines a different potential. Let us now recall that the corresponding expression for the transformed potential was obtained in \cite{rosu1,rosu2}. 
Adopting the notation used in the latter reference, this potential reads
\begin{eqnarray}
V_{1 \gamma} &=& V_1 -2~\frac{d^2}{dx^2}~ \log\left(
\gamma+\int\limits^x \Psi_0^2~dt
\right), \label{rosupot}
\end{eqnarray}
where  $\gamma$ is an arbitrary constant and $\Psi_0$ solves (\ref{eq1}) for 
$\epsilon=0$ (zero-energy solution). If we now compare (\ref{v3}) and (\ref{rosupot}), we observe that $V_{1\gamma}$ for 
$\gamma= c_1/c_2$ is a 
special case of $V_3$. While $\Psi_0$ can be interpreted as a transformation function of our first SUSY transformation (\ref{susy1}) for 
the factorization energy $\lambda=0$, its counterpart $u$ in (\ref{v3}) does not carry any constraints regarding its factorization 
energy. In particular, if our initial Schr\"odinger equation admits a discrete spectrum, $\Psi_0$ represents the ground state 
solution, while $u$ can stand for the ground state, an excited state, or any other solution of (\ref{aux1}). Let us now 
construct a solution to the Schr\"odinger equation (\ref{eq3}). To this end, we must distinguish the 
cases $\epsilon=\lambda$ and $\epsilon \neq \lambda$. In the first case $\epsilon=\lambda$ we know that 
$\hat{\Psi}_p=1/\bar{u}$ is 
a particular solution of (\ref{eq3}). Upon application of the reduction-of-order formula we obtain its general 
solution in the form
\begin{eqnarray}
\hat{\Psi} &=& \hat{C}_1~\hat{\Psi}_p+\hat{C}_2~\hat{\Psi}_p~\int\limits^x \frac{1}{\hat{\Psi}^2_p}~dt \nonumber \\
&=&  \frac{\hat{C}_1}{\bar{u}}+\frac{\hat{C}_2}{\bar{u}}~\int\limits^x\bar{u}^2~dt \nonumber \\
&=& \frac{u}{\frac{c_1}{c_2} +~\int\limits^x u^2~dt} \left[C_1+C_2~
\int\limits^x\frac{1}{u^2} \left(\frac{c_1}{c_2}+\int\limits^t u^2~ds\right)^2 ~dt
\right]. \label{solzero}
\end{eqnarray}
where $C_1$ and $C_2$ are arbitrary constants. Since every ratio $c_1/c_2$ defines a potential \eqref{v3} and as a consequence a Schr\"odinger equation \eqref{eq3}, this parameter is also found in its solutions \cite{Rosu98}. We will now compare this expression to the corresponding result that was given in \cite{rosu1,rosu2}. It reads
\begin{eqnarray}
\Psi_{0 \gamma} &=& \frac{\Psi_0}{\gamma+\int\limits^x \Psi_0^2~dt}. \label{solrosu}
\end{eqnarray}
It is straightforward to see that (\ref{solrosu}) is a special case of (\ref{solzero}) if we employ the settings $c_1/c_2= \gamma, ~C_1 = 1, ~C_2=0$ and further identify $u=\Psi_0$. In the context of spectral 
problems, (\ref{solrosu}) is known as the ``missing state'' solution, because its particular form 
can be associated with a spectral value that has possibly been removed or added by the SUSY transformation. In general, 
all solutions resulting from a confluent SUSY transformation, can be expressed in terms of Wronskians, in 
particular for higher-order cases \cite{xbatconfluent}. As an example, let us mention the Wronskian representation 
for the solution (\ref{solrosu}):
\begin{eqnarray}
\Psi_{0 \gamma} &=& \frac{W_{\Psi_0,u_1,v_0}}{W_{\Psi_0,u_1}}, \label{wrep}
\end{eqnarray}
where $v_0$ and $\Psi_0$ solve the same equation with $W_{v_0,\Psi_0}=1$. The function $u_1$ is a solution of 
the inhomogeneous equation
\begin{eqnarray}
u_1''-V_1~u_1 &=& -\Psi_0. \nonumber
\end{eqnarray}
Evaluation of (\ref{wrep}) leads to the identity \cite{fernandezreview}
\begin{eqnarray}
\Psi_{0 \gamma} &=& \frac{\Psi_0}{W_{\Psi_0,u_1}} ~=~ \frac{\Psi_0}{\gamma+\int\limits^x \Psi_0^2~dt}, \nonumber
\end{eqnarray}
which coincides with (\ref{solrosu}) for a constant of integration $\gamma$. 

In the second case, when $\epsilon \neq \lambda$, the solutions $\Psi$ of equation \eqref{eq1} are transformed as follows:
\begin{eqnarray}
\hat{\Psi} = \frac{W_{u,u_1,\Psi}}{W_{u,u_1}}, 
\end{eqnarray}
where $u_1$ satisfies the inhomogeneous equation 
\begin{eqnarray}
u_1''+(\lambda -V_1)~u_1 &=& -u. \nonumber
\end{eqnarray}
The reader might refer to \cite{dj12,xbatconfluent,bagrov} for further information about the general concept of Wronskian representation. Let us now summarize the benefits of embedding the zero-mode SUSY scheme into the context of the confluent SUSY algorithm. The principal advantage is that the SUSY confluent algorithm provides the possibility of constructing 
quantum potentials  using a transformation function $u$ shown in (\ref{v3}) that does not need to be the zero-energy mode of the initial Schr\"odinger equation 
(\ref{eq1}), but can be any solution of the latter equation. In spectral problems, this feature can be used for 
creating and annihilating bound states that lie above the ground state level, and also to generate isospectral systems \cite{baye2,baye1,contreras08,Midya11}. 
Furthermore, it is straightforward to 
extend the scheme used in the latter two references to the case of higher-order SUSY transformations, which can be either 
confluent or a mixture between confluent and conventional transformations. Such scenarios have been studied in 
\cite{xbatconfluent}.

\section{Applications} \label{Section application}
We will now present applications of the generalized SUSY scheme that was introduced in section 2. The focus of our 
applications will be twofold. We show how our scheme can be used for models not governed by a Schr\"odinger equation, 
and we choose the transformation function $u$ to solve (\ref{aux1}) for $\lambda \neq 0$. In the following applications we will adopt the the notation $\gamma = c_1/c_2$ in \eqref{v3} and \eqref{solzero} as a parameter that characterize the transformation.  

\subsection{Dirac equation for an inverted harmonic oscillator potential}
\begin{figure}[h]
\begin{center}
\epsfig{file=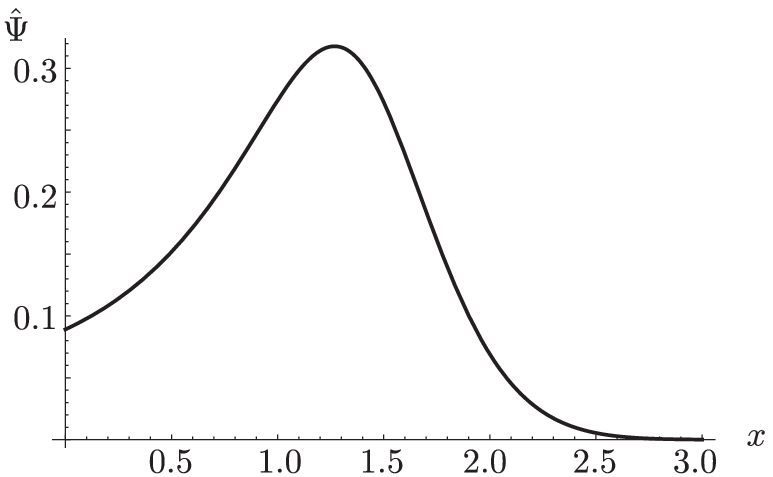,width=7.9cm}
\epsfig{file=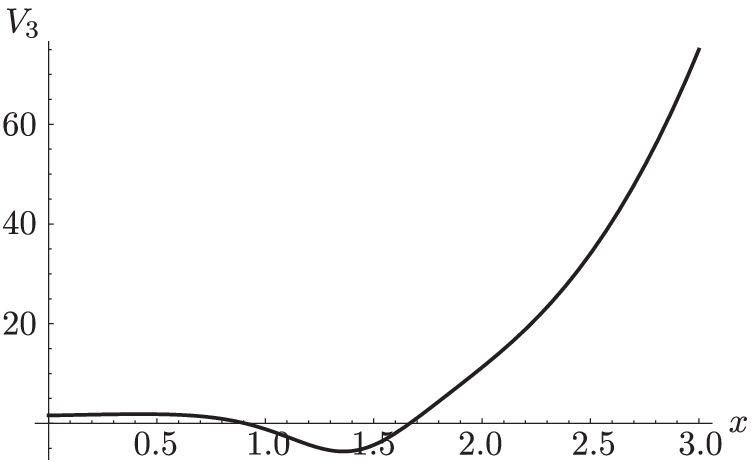,width=7.9cm}
\caption{The solution (\ref{hatpsiex}) in the left plot and the corresponding potential (\ref{v3ex}) in the right plot.
} 
\label{fignew}
\end{center}
\end{figure}

In this example we consider the following stationary, two-component Dirac equation in 
one spatial dimension 
\begin{eqnarray}
i~\sigma_2~\Phi' + (V_0 - E)~\Phi &=& 0, \label{dirac}
\end{eqnarray}
where $\sigma_2$ stands for the second Pauli matrix, $E$ denotes a real-valued number, and 
the spinor $\Phi=(\Phi_1,\Phi_2)^T$ represents the solution. Furthermore, we assume the potential $V_0$ 
to be of pseudoscalar form, that is, 
\begin{eqnarray}
V_0 &=& m~\sigma_3+q_0~\sigma_1, \label{pseudo}
\end{eqnarray}
for a constant, positive mass $m$, Pauli matrices $\sigma_1$, $\sigma_3$, and a function $q_0$ that is 
given by
\begin{eqnarray}
q_0 &=& -x^2. \label{q0}
\end{eqnarray}
This function can be interpreted as an inverted oscillator potential. In order to construct a solution to the Dirac 
equation (\ref{dirac}), we write it in components
\begin{eqnarray}
\Phi_1'+x^2~\Phi_1+ \left( E+m\right) \Phi_2 &=& 0, \label{d1} \\
\Phi_2'-x^2~\Phi_2- \left( E-m \right) \Phi_1 &=& 0, \label{d2}
\end{eqnarray}
where we took the form (\ref{pseudo}) of the potential and its parametrizing function (\ref{q0}) into account. The 
system (\ref{d1}), (\ref{d2}) does not admit closed-form solutions except in the case $|E|=m$, which we will focus on 
from now on. We find the general solution of the system to be
\begin{eqnarray}
\Phi_1 &=& k_1~\exp\left(\int\limits^x q_0~dt \right)-2~m~\exp\left(\int\limits^x q_0~dt \right) 
\int\limits^x \Phi_2~\exp\left(\int\limits^t -q_0~ds \right) dt \label{d1sol} \\[1ex]
\Phi_2 &=& k_2~\exp\left(-\int\limits^x q_0~dt \right), \label{d2sol}
\end{eqnarray}
introducing two arbitrary constants $k_1$ and $k_2$. On substituting (\ref{q0}), this solution simplifies to
\begin{eqnarray}
\Phi_1 &=& k_1~\exp\left(-\frac{1}{3}~x^3 \right)+\frac{2}{3}~k_2~m~\exp\left(-\frac{1}{3}~x^3 \right) x
~E_{\frac{2}{3}}\left(-\frac{2}{3}~x^3 \right) \label{phi1} \\[1ex]
\Phi_2 &=& k_2~\exp\left(-\frac{1}{3}~x^3 \right), \nonumber
\end{eqnarray}
where $E$ stands for the exponential integral function \cite{abram}. 
We will now apply the SUSY transformation scheme that was introduced in section 2 \cite{contreras14}. Since the latter scheme 
starts from a second-order equation of Schr\"odinger type, we must first convert the system (\ref{d1}), (\ref{d2}) 
into such an equation. Solving (\ref{d1}) for 
$\Phi_2$ and substitution of the resulting expression into (\ref{d2}) leads to 
\begin{eqnarray}
\Phi_1''+\left(-x^4+2~x\right) \Phi_1 &=& 0. \label{schr}
\end{eqnarray}
Let us note that without incorporation of the explicit form (\ref{q0}), the latter equation would have the form
\begin{eqnarray}
\Phi_1''+\left(q_0^2+q_0'\right) \Phi_1 &=& 0. \label{schrq0}
\end{eqnarray}
This will become important below, when we need to reverse the steps of passing from the Dirac equation to its 
Schr\"odinger counterpart. The second-order equation (\ref{schr}) can be compared with (\ref{eq1}), yielding the identifications
\begin{eqnarray}
\Psi ~=~ \Phi_1 \qquad \qquad \epsilon~=~0 \qquad \qquad V_0 ~=~ x^4-2~x, \nonumber
\end{eqnarray}
recall that $\Phi_1$ is given in (\ref{phi1}). In order to perform the SUSY transformation (\ref{solzero}), we must 
provide a transformation function $u$ that is a solution of equation (\ref{schr}). As such, $u$ must have the form 
(\ref{phi1}) or it can be any special case of it for a particular choice of parameters $k_1$ and $k_2$. Due to the 
length of the expressions involved we omit to state the full form of the transformed function (\ref{solzero}). Instead, 
we employ the following parameter setting
\begin{eqnarray}
k_1~=~1, \qquad C_1~=~-1/10, \qquad \gamma= -1/10, \qquad k_2~=~C_2~=~0.  \label{parset}
\end{eqnarray}
For these parameters, the function $\hat{\Psi}$ takes the form
\begin{eqnarray}
\hat{\Psi} &=& \frac{3~\exp\left(-\frac{1}{3}~x^3 \right)}{3+10~x~\mbox{E}_{\frac{2}{3}}\left(\frac{2}{3}~x^3 \right)}, 
\label{hatpsiex}
\end{eqnarray}
note that E represents the exponential integral function \cite{abram}. 
The corresponding potential $V_3$ that is given in (\ref{v3}) reads for the present case
\begin{eqnarray}
V_3 &=& x^4-2~x-\frac{120~x^2~\exp\left(-\frac{2}{3}~x^3 \right)}{
3+10~x~\mbox{E}_{\frac{2}{3}}\left(\frac{2}{3}~x^3 \right)}+
\frac{1800~\exp\left(-\frac{4}{3}~x^3 \right)}{\left[
3+10~x~\mbox{E}_{\frac{2}{3}}\left(\frac{2}{3}~x^3 \right)\right]^2}
. \label{v3ex}
\end{eqnarray}
The solution (\ref{hatpsiex}) and its potential (\ref{v3ex}) are displayed in figure \ref{fignew}. Note that both solution and potential are bounded at the origin and nonsingular on the positive real semi-axis. 
As in the case of the function $\hat{\Psi}$, the general expression for $V_3$ is too large to be displayed here. 

\begin{figure}[h]
\begin{center}
\epsfig{file=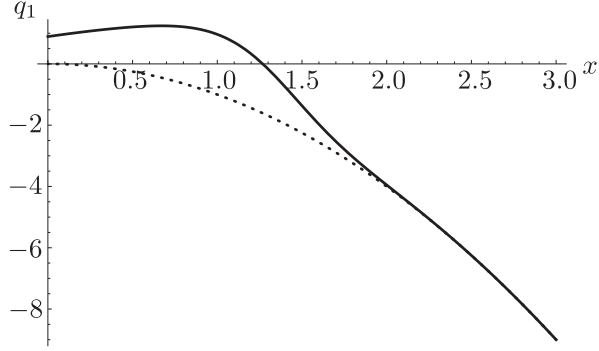,width=8cm}
 \caption{The parametrizing function $q_0$ (dashed curve), together with its transformed counterpart $q_1$ for the settings 
$k_1=1,~\gamma=-0.1,~C_2=0$ (solid curve).
} 
\label{figq1}
\end{center}
\end{figure}

Now that we have completed the confluent SUSY scheme, we must relate (\ref{hatpsiex}) and 
(\ref{v3ex}) with a transformed Dirac equation that has the form (\ref{dirac}). To this end, we first recall that 
$\hat{\Psi}$ is a solution to (\ref{eq3}) for the potential $V_3$. If we can rewrite the latter potential as 
\begin{eqnarray}
V_3 &=& q_1^2+q_1', \nonumber
\end{eqnarray}
then $q_1$ will be the parametrizing function of the transformed pseudoscalar potential in a Dirac equation 
of the form (\ref{dirac}). It is well-known that $q_1$ can be constructed by means of the setting
\begin{eqnarray}
q_1 &=& \frac{\hat{\Psi}'}{\hat{\Psi}}. \nonumber
\end{eqnarray}		
Note that the function $\hat{\Psi}$ was determined, but we only displayed the particular case (\ref{hatpsiex}) here. 
Following this idea, we show the explicit form of $q_1$ for the setting $k_2=0$
\begin{eqnarray}
q_1 &=& -x^2+\frac{30~\exp\left(-\frac{2}{3}~x^3 \right)}{3+10~x~\mbox{E}_{\frac{2}{3}}\left(\frac{2}{3}~x^3 \right)}. \label{q1}
\end{eqnarray}
Figure \ref{figq1} shows a particular case of the function $q_1$, together with its initial partner function $q_0$.

It remains to find the solution of the Dirac equation for the pseudoscalar potential parametrized by the function $q_1$, that is,
\begin{eqnarray}
i~\sigma_2~\hat{\Phi}' + (V_1 - E)~\hat{\Phi} ~=~ 0, ~~~~~~V_1 ~=~ m~\sigma_3+q_1~\sigma_1, \label{diract}
\end{eqnarray}
where we assume that $|E|=m$. Since the latter Dirac equation has the same form as (\ref{dirac}), its solutions must be of the form 
(\ref{d1sol}), (\ref{d2sol}), where $q_0$ is to be replaced by $q_1$, as shown in (\ref{q1}). Since the integrals in the 
resulting expression cannot be resolved, we omit to show the combination of (\ref{q1}) and (\ref{d1sol}), (\ref{d2sol}). 
Hence, the application of the SUSY scheme to the pair of Dirac equations (\ref{dirac}), (\ref{diract}) is complete.

\subsection{Schr\"odinger equation for a constant potential}

We will now pick up an example that was presented in \cite{rosu2} and show that we can generalize it considerably when 
using the full scheme introduced in section 2. Let us consider the Schr\"odinger equation (\ref{eq1}) for the setting 
$V_1=c^2$, where $c$ is an arbitrary real number:
\begin{eqnarray}
\Psi''+(\epsilon-c^2)~\Psi &=& 0. \label{eq1con}
\end{eqnarray}
This equation has the following general solution
\begin{eqnarray}
\Psi &=& k_1~\exp\left(\sqrt{c^2-\epsilon}~x \right)+k_2~\exp\left(-\sqrt{c^2-\epsilon}~x \right), \label{solcon}
\end{eqnarray}
where $k_1$ and $k_2$ stand for arbitrary constants. In the work \cite{rosu2}, the following parameter setting was 
considered
\begin{eqnarray}
u ~=~ \exp\left(-c~x \right), \qquad
\epsilon ~=~ 0, \qquad   C_1~=~1, \qquad  C_2~=~0. \label{rosuset}
\end{eqnarray}
Here, the function $u$ is a special case of (\ref{solcon}) for zero energy. 
As a result of (\ref{rosuset}), the following transformed potential (\ref{v3})  and corresponding solution (\ref{solzero})
were obtained 
\begin{eqnarray}
V_3 &=& c^2+\frac{8~c^2}{2~c~\gamma~\exp\left(2~c~x \right)-1}+\frac{8~c^2}{\left[2~c~\gamma~\exp\left(2~c~x \right)-1 \right]^2}, 
\label{v3rosu} \\[1ex]
\hat{\Psi} &=& \frac{2~c~\exp\left(c~x \right)}{2~c~\gamma\exp\left(2~c~x \right)-1}. \label{solrosucon}
\end{eqnarray}
This potential and the zero-energy solution coincide exactly with the finding in \cite{rosu2}, except for some minor 
algebraic simplifications. We are now ready to construct a generalized version of both (\ref{v3rosu}) and (\ref{solrosucon}). 
While the approach presented in 
\cite{rosu2} requires the function $u$ in (\ref{rosuset}) to be a zero-energy solution of the initial equation (\ref{eq1con}), 
our scheme allows to drop this requirement. We will now use the complete solution (\ref{solcon}) to evaluate (\ref{v3}) and (\ref{solzero}):
\begin{eqnarray}
V_3 &= &  c^2  - \frac{4(c^2-\epsilon) \left[ k_1^2 \exp \left( 2 \sqrt{c^2-\epsilon} ~x\right)- k_2^2  \exp\left( -2 \sqrt{c^2-\epsilon}~ x\right)\right]}{  2 \gamma  \sqrt{c^2-\epsilon} +k_1^2 \exp\left(2 \sqrt{c^2-\epsilon}~ x\right)+4 k_1 k_2 \sqrt{c^2-\epsilon}~ x-k_2^2 \exp\left(-2 \sqrt{c^2-\epsilon} ~x\right)}  \nonumber\\
&+&  \frac{ 4 (c^2-\epsilon) \left[ k_1^2  \exp\left(2 \sqrt{c^2-\epsilon} ~x\right)+ k_2^2  \exp\left(-2 \sqrt{c^2-\epsilon}~ x\right)+2 k_1 k_2 \right]^2}{ \left[2 \gamma  \sqrt{c^2-\epsilon} +k_1^2 \exp \left(2 \sqrt{c^2-\epsilon}~ x\right)-k_2^2 \exp \left(-2 \sqrt{c^2-\epsilon}~ x \right) +4 k_1 k_2 \sqrt{c^2-\epsilon} ~x\right]^2} \label{V3gen}
\end{eqnarray}
and 
\begin{align}
 \hat{\Psi}& =  \frac{2 \sqrt{c^2-\epsilon} \left[k_1~\exp\left(\sqrt{c^2-\epsilon}~x \right)+k_2~\exp\left(-\sqrt{c^2-\epsilon}~x \right)\right]}{\left[2 \gamma  \sqrt{c^2-\epsilon} +k_1^2 \exp \left(2 \sqrt{c^2-\epsilon} ~x\right)-k_2^2 \exp \left(-2 \sqrt{c^2-\epsilon} ~x \right) +4 k_1 k_2 \sqrt{c^2-\epsilon} ~x \right]}  \nonumber \\
~ & \times \left\{ C_1 + C_2 \left( \frac{1}{2  \sqrt{c^2-\epsilon} }\right)^3 \left[ k_1^2 \exp \left(2 \sqrt{c^2-\epsilon} ~x\right)-k_2^2 \exp \left(-2 \sqrt{c^2-\epsilon} ~x \right)-8 k_1 k_2(c^2-\epsilon)~ x^2 \right. \right. \nonumber  \\
 & \left. \left. - 4  \sqrt{c^2-\epsilon} \left(k_1 k_2 +2 \gamma \sqrt{c^2-\epsilon} \right) x   + \frac{4(c^2-\epsilon)(2 k_1 k_2 x +\gamma)^2 }{k_1 k_2+k_2^2 \exp\left(-\sqrt{c^2-\epsilon}~x \right)   }    \right] \right\}, \label{PSI}
\end{align}
note that we leave the energy $\epsilon$ and the constants $k_1,~ k_2$ undetermined. The last two expressions are very general and some particular values for the parameters involved can considerably reduce the expressions or be interesting in the physical point of view. 

\begin{figure}[h]
\begin{center}
\epsfig{file=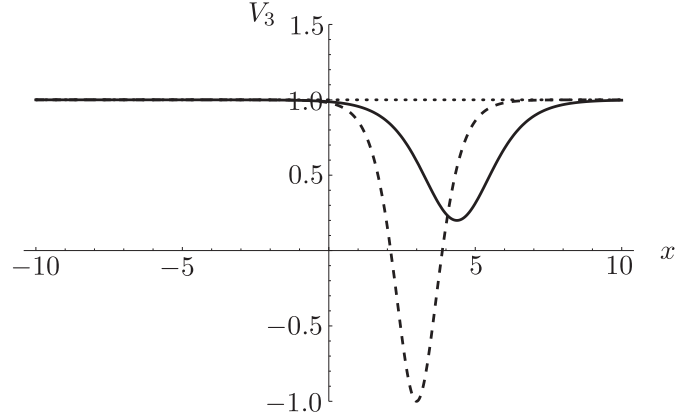,width=9cm}
\caption{The constant potential $V_1=1$ (dotted curve) and the potentials $V_3$ with the parameters $\epsilon=0, ~\gamma=200$ (dashed curve) and $\epsilon =0.6, ~\gamma = 200$ (solid curve).
} 
\label{constant}
\end{center}
\end{figure}

\paragraph{Potentials with one bound state}
In equation \eqref{v3rosu} can be seen a family of potentials characterized by the parameter $\gamma$ and \eqref{solrosucon} presents its only bound state at zero energy. Now let us consider the following setting 
\begin{eqnarray}
u ~=~ \exp\left(\sqrt{c^2-\epsilon}~x \right), \qquad \epsilon < c^2 \qquad,  C_1~=~1, \qquad  C_2~=~0 . \label{set1}
\end{eqnarray}
The potentials obtained after the confluent SUSY transformation and its bound state are 
\begin{eqnarray}
V_3 &= & c^2- \frac{8 (c^2-\epsilon)}{2 \gamma \sqrt{c^2-\epsilon} \exp(-2\sqrt{c^2-\epsilon}~x)+1}+\frac{8 (c^2-\epsilon)}{\left[ 2 \gamma \sqrt{c^2-\epsilon} \exp(-2\sqrt{c^2-\epsilon}~x)+1\right]^2}~, \\
\hat{\Psi}&=& \frac{2  \sqrt{c^2-\epsilon} \exp(\sqrt{c^2-\epsilon}~x) }{2 \gamma \sqrt{c^2-\epsilon} +\exp(2\sqrt{c^2-\epsilon}~x)  }. \label{conbound}
\end{eqnarray}
In order to avoid singularities in the new potential $\gamma$ has to be a positive constant. The function \eqref{conbound} is the only bound state of the system. Figure \ref{constant} shows the graph of the potential $V_1=c^2$ (dotted curve) and two potentials $V_3$ with different parameters. In the first case we use a zero-mode solution as transformation function (dashed curve) and then a solution of \eqref{aux1} where $\lambda = 0.6$ (solid curve). In both cases the systems have a bound state associated to $\epsilon = 0$ and $\epsilon = 0.6$, respectively. 

\paragraph{Hyperbolic case}
Simplification of \eqref{V3gen} and \eqref{PSI} considering $k_1 = k_2= C_1=1, ~ C_2=0$ and also $\epsilon < c^2$ gives the following results 
\begin{eqnarray}
V_3 &=& c^2 - \frac{16 (c^2-\epsilon) \cosh(\sqrt{c^2-\epsilon}~x) \sinh (\sqrt{c^2-\epsilon}~x)}{\sqrt{c^2-\epsilon}(\gamma+2x)+\sinh (2 \sqrt{c^2-\epsilon} ~ x)}  \nonumber \\
&+& \frac{16 (c^2-\epsilon) \cosh^4 (\sqrt{c^2-\epsilon}~x)}{\left[ \sqrt{c^2-\epsilon}(\gamma+2x)+\sinh (2 \sqrt{c^2-\epsilon} ~ x) \right]^2}~, \\
\hat{\Psi}&=& \frac{2  \sqrt{c^2-\epsilon} \cosh( \sqrt{c^2-\epsilon} ~x) }{ \sqrt{c^2-\epsilon}(\gamma +2x)+\sinh(2  \sqrt{c^2-\epsilon} ~x)}~,
\end{eqnarray}
where for the sake of simplicity we introduced hyperbolic functions. The family of potentials is singular for all the values of $\gamma$ and as a consequence no bound states can be found.

\paragraph{Trigonometric case}
When the parameter $\epsilon$ is greater than the potential $V_1=c^2$ the exponential functions  in \eqref{V3gen} and \eqref{PSI} can be expressed as trigonometric functions. With the following setting
\begin{eqnarray}
u ~=~ \sin\left(\sqrt{c^2-\epsilon}~x \right), \qquad \epsilon > c^2,  \qquad  C_1~=~1, \qquad  C_2~=~0, \label{set1}
\end{eqnarray}
the transformed potential and the corresponding solution of the Schr\"odinger equation are 

\begin{align}
V_3=& c^2 - \frac{8(c^2-\epsilon) \sin(2\sqrt{c^2-\epsilon}~x)}{2 \sqrt{c^2-\epsilon} (2\gamma +x )-\sin(2\sqrt{c^2-\epsilon}~x)}+\frac{32(c^2-\epsilon) \sin^4 (\sqrt{c^2-\epsilon}~x)}{\left[2 \sqrt{c^2-\epsilon} (2\gamma +x )-\sin(2\sqrt{c^2-\epsilon}~x)  \right]^2}~, \\
\hat{\Psi}=&\frac{4\sqrt{c^2-\epsilon}~\sin(\sqrt{c^2-\epsilon}~x)}{2 \sqrt{c^2-\epsilon} (2\gamma +x )-\sin(2\sqrt{c^2-\epsilon}~x)}.
\end{align}
If we consider the whole real axis as domain then there are no regular potentials $V_3$ obtained  with the confluent SUSY transformation.

\subsection{Fokker-Planck equation with constant diffusion coefficient}
\begin{figure}[h]
\begin{center}
\epsfig{file=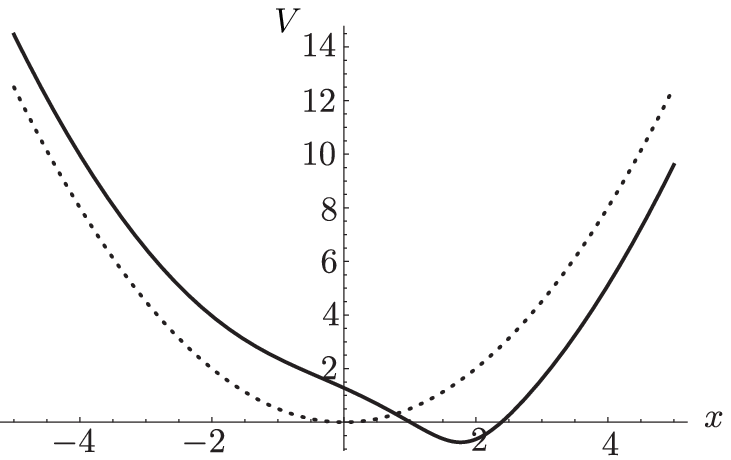,width=7.9cm}
\epsfig{file=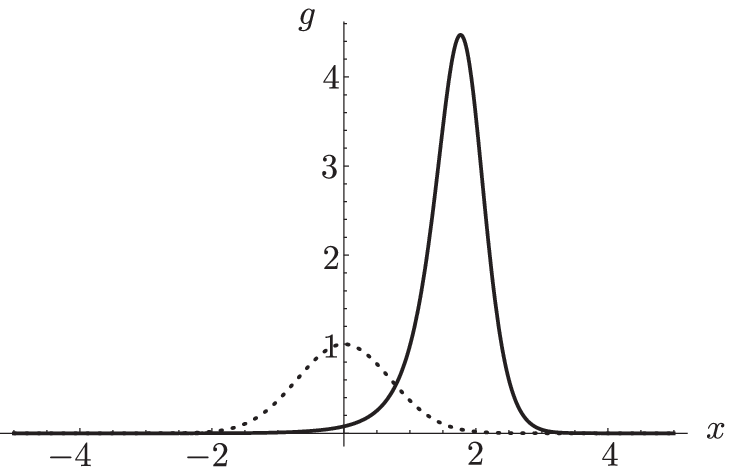,width=7.9cm}
\caption{Left plot: the transformed drift function (\ref{v0}) (solid curve), together with its initial counterpart (\ref{drift}) 
(dashed curve) for the 
settings (\ref{set0}). Right plot: solutions of the Fokker-Planck equations (\ref{fokkert}) (solid curve) 
for $t=0$ and (\ref{bvp1}) (dashed curve).
} 
\label{figv0}
\end{center}
\end{figure}

The SUSY scheme introduced in section 2 can also be applied to certain cases of the Fokker-Planck equation, as will be 
shown in this subsection \cite{Rosu97,Schulze09}. Let us consider the following equation
\begin{eqnarray}
\frac{\partial f}{\partial t}-\frac{1}{2}~\frac{\partial^2 f}{\partial x^2}-\frac{\partial}{\partial x} \left(U' f \right) &=& 0, 
~~~(x,t) \in (0,\infty) \times (0,\infty). \label{bvp1}
\end{eqnarray}
This Fokker-Planck equation (\ref{bvp1}) has a constant diffusion coefficient and a drift function $U$ that we assume 
to depend on the spatial variable only. We will comment on boundary and initial conditions once we have constructed 
the transformed equation and its solutions. In order to apply our SUSY algorithm, we must convert (\ref{bvp1}) into a Schr\"odinger equation of the form (\ref{eq1}). 
To this end, we rewrite the solution $f$ as follows:
\begin{eqnarray}
f &=& \exp\left(-U-k~t\right)~\Psi, \label{f}
\end{eqnarray}
where $k$ is a real constant and $\Psi$ is a function independent of the time variable. Substitution of (\ref{f}) into equation 
(\ref{bvp1}) and simplification gives the following result
\begin{eqnarray}
\Psi''+\left[2~k-(U')^2+U'' \right] \Psi &=& 0, \label{schrfok}
\end{eqnarray}
This matches our Schr\"odinger equation (\ref{eq1}) if the following settings are employed
\begin{eqnarray}
\epsilon ~=~ 2~k \qquad \qquad \qquad V_1 ~=~ (U')^2-U''. \nonumber
\end{eqnarray}
In the next step we specify the drift function $U$ in (\ref{bvp1}) as follows:
\begin{eqnarray}
U &=& \frac{x^2}{2}. \label{drift}
\end{eqnarray}
Let us now rewrite equation (\ref{schrfok}) using the explicit form (\ref{drift}) of $U$:
\begin{eqnarray}
\Psi''+\left[2~k-x^2+1 \right] \Psi &=& 0. \label{schrfokex}
\end{eqnarray}
The general solution of this equation is given by
\begin{eqnarray}
\Psi &=& k_1~\exp\left(-\frac{x^2}{2} \right)~H_k(x) + k_2~\exp\left(\frac{x^2}{2} \right)~H_{-k-1}(i~x), \label{gensol}
\end{eqnarray}
where $k_1, k_2$ are free constants and $H$ stands for the Hermite function \cite{abram}. Recall that for an index that is a 
nonnegative integer, the Hermite function becomes a polynomial. In the next step, we must choose a solution $u$ of 
(\ref{schrfokex}) that enters in both the potential (\ref{v3}) and the corresponding function (\ref{solzero}). 
We pick the following special case of (\ref{gensol})
\begin{eqnarray}
u &=& \exp\left(-\frac{x^2}{2} \right)~H_k(x), \label{ufok}
\end{eqnarray}
where $k$ is a nonnegative integer. Substitution of (\ref{ufok}) into (\ref{v3}) yields
\begin{eqnarray}
V_3 &=& x^2-1+\frac{2~\exp\left(-2~x^2 \right)~H_k^4(x)}{\left[
\gamma+\int\limits^x \exp\left(-t^2 \right)~H_k^2(t)~dt
\right]^2} \nonumber \\[1ex] 
&+& \frac{4~\exp\left(-x^2 \right)~x~H_k^2(x)-8~k~\exp\left(-x^2 \right)~H_{k-1}(x)~H_k(x)
}{
\gamma+\int\limits^x \exp\left(-t^2 \right)~H_k^2(t)~dt
}. \label{v3fok}
\end{eqnarray}
The solution of our Schr\"odinger equation (\ref{eq3}) can be found by inserting (\ref{ufok}) into (\ref{solzero}). We get
\begin{eqnarray}
\hat{\Psi} = \frac{\exp\left(-\frac{x^2}{2} \right)~H_k(x)}{\gamma+~\int\limits^x \exp\left(-t^2 \right) H_k^2(t)~dt}
\left\{C_1+C_2 \hspace{-.1cm} \int\limits^x \frac{\exp\left(-t^2 \right)}{H_k^2(t)}
\left[\gamma+\int\limits^t \exp\left(-s^2 \right) H_k^2(s)~ds
\right]^2 \hspace{-.2cm} dt
\right\}. \nonumber  
\end{eqnarray}
\vspace{-.5cm}
\begin{eqnarray}
\label{solzerofok}
\end{eqnarray}
In the last step we need to convert (\ref{solzerofok}) into the solution of a Fokker-Planck equation, the drift function of which is 
determined by (\ref{v3fok}). To this end, let us introduce the following quantity
\begin{eqnarray}
V &=& -\log\left(\hat{\Psi}\right), \label{driftt}
\end{eqnarray}
where $\hat{\Psi}$ is defined in (\ref{solzerofok}). It is then straightforward to verify that the Schr\"odinger equation (\ref{eq3}) 
for the potential (\ref{v3fok}) can be rewritten as follows
\begin{eqnarray}
\hat{\Psi}''+\left[-(V')^2+V''
\right] \hat{\Psi} &=& 0. \label{eq4}
\end{eqnarray}
Note that for the sake of clarity we did not include the full form of $V$, which can be obtained from combining 
(\ref{solzerofok}) and (\ref{driftt}). Now, comparison of (\ref{eq4}) with (\ref{schrfok}) with (\ref{driftt})
shows that the function
\begin{eqnarray}
g &=& \exp\left(-V \right) \hat{\Psi} ~=~ \hat{\Psi}^2, \label{g}
\end{eqnarray}
is a time-independent solution of the Fokker-Planck equation
\begin{eqnarray}
\frac{\partial g}{\partial t}-\frac{1}{2}~\frac{\partial^2 g}{\partial x^2}-\frac{\partial}{\partial x} \left(V' g \right) &=& 0. 
\label{fokkert}
\end{eqnarray}
We observe that (\ref{driftt}) is the transformed drift function, while the corresponding solution (\ref{g}) is obtained by 
squaring $\hat{\Psi}$ in (\ref{solzerofok}). Since the integrals in the latter expression cannot be resolved in their general 
form, we will now restrict ourselves to a particular case. Let us employ the settings
\begin{eqnarray}
\gamma=-0.9, \qquad \qquad C_1~=~-0.25 \qquad \qquad C_2~=~0 \qquad \qquad k~=~0. \label{set0}
\end{eqnarray}
This gives the transformed drift function as 
\begin{eqnarray}
V &=& -\log\left[\frac{
5~\exp\left(-\frac{x^2}{2} \right)}
{18-10~\sqrt{\pi}~\mbox{erf}(x)}
\right], \label{v0}
\end{eqnarray}
where the symbol erf stands for the error function \cite{abram}. The left plot in figure \ref{figv0} shows the graphs of the 
drift functions $V$ and $U$, as given in (\ref{v0}) and (\ref{drift}), respectively. 

Inspection of expression (\ref{v0}) shows that the transformed drift function $V$ can be defined on the whole 
real line. The same holds for the associated solution of the transformed Fokker-Planck equation (\ref{fokkert}), 
see the right plot in figure \ref{figv0}. In addition, we observe that this function vanishes at the infinities.A study of expression \eqref{driftt} indicates that for $\gamma \in (-\infty,0)$ when $k=0$ and $C_1 < 0$ will lead us to a regular drift function $V$.

\begin{figure}[h]
\begin{center}
\epsfig{file=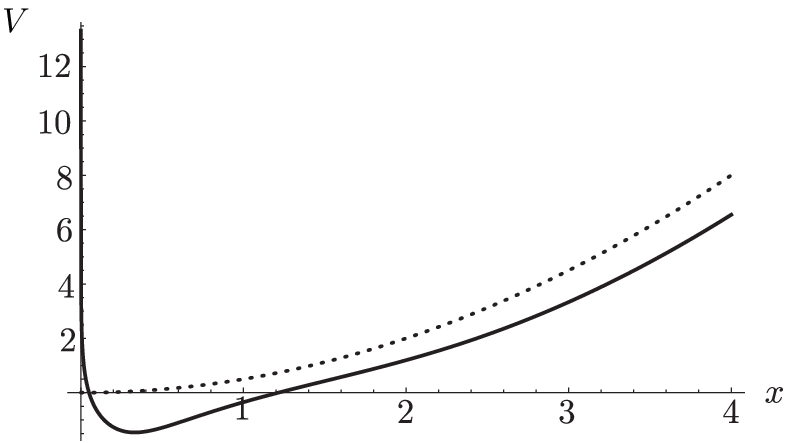,width=7.9cm}
\epsfig{file=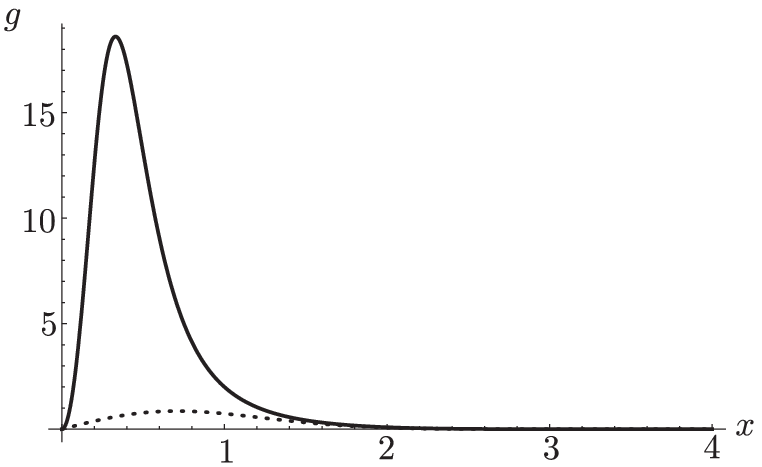,width=7.9cm}
\caption{Left plot: the transformed drift function (\ref{v1}) (solid curve), together with its initial counterpart (\ref{drift}) 
(dashed curve) for the 
settings (\ref{set1}). Right plot: solutions of the Fokker-Planck equations (\ref{fokkert}) (solid curve) 
for $t=0$ and (\ref{bvp1}) (dashed curve).
} 
\label{figv1}
\end{center}
\end{figure}

Let us now 
choose a set of parameters that is different from (\ref{set0})
\begin{eqnarray}
\gamma=0.1, \qquad \qquad C_1~=~1, \qquad \qquad C_2~=~0, \qquad \qquad k~=~1. \label{set1}
\end{eqnarray}
For these settings, the transformed drift function reads
\begin{eqnarray}
V &=& -\log\left[\frac{
20~x~\exp\left(\frac{x^2}{2} \right)}
{-20~x+\exp\left(x^2 \right) \left[1+10~\sqrt{\pi}~\mbox{erf}(x) \right]
}
\right]. \label{v1}
\end{eqnarray}
Similar to the previous case, the left and right plots in figure \ref{figv1} show the drift functions (\ref{drift}), (\ref{v1}) and 
the associated solutions of (\ref{bvp1}), (\ref{fokkert}), respectively. We observe that the transformed solutions can be 
defined on the positive half-axis, vanishing 
at zero and at positive infinity. Furthermore, a study of equation \eqref{driftt} under the set of parameters $ C_1=1,~ C_2=0,~ k=1$ shows that when $\gamma \in (0,\infty)$ the only singularity of the drift function $V$ is in the origin. 

\begin{figure}[h]
\begin{center}
\epsfig{file=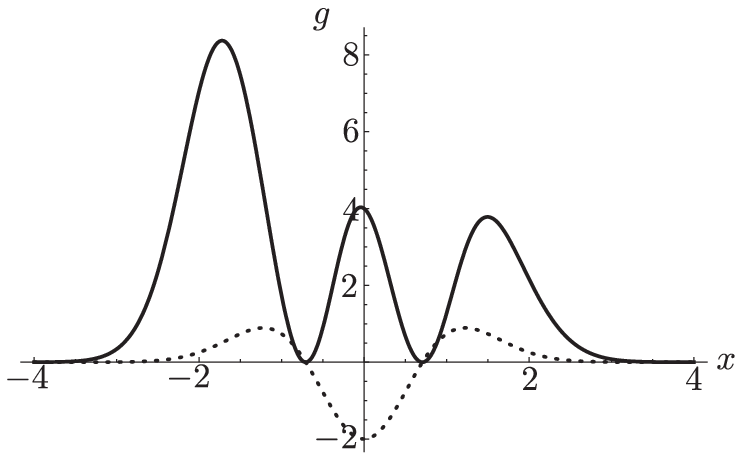,width=7.9cm}
\epsfig{file=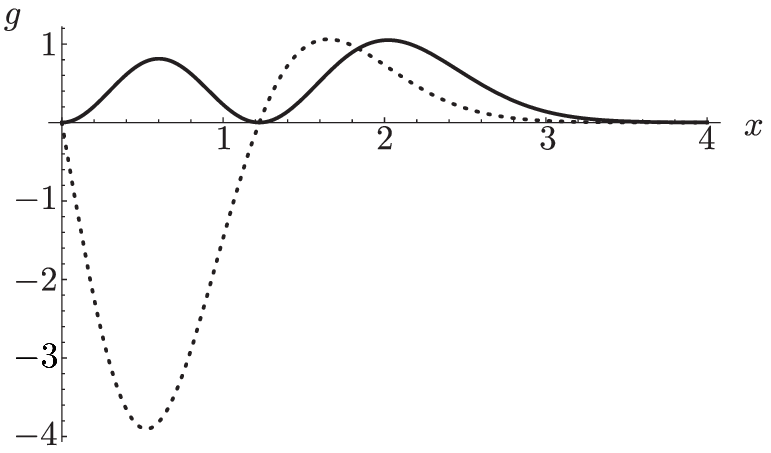,width=7.9cm}
\caption{Solutions of the Fokker-Planck equations (\ref{fokkert}) (solid curve) 
for $t=0$ and (\ref{bvp1}) (dashed curve) for the 
settings $C_1=20,~C_2=0,~k=2,~\gamma=20$ (left plot) and $C_1=20,~C_2=0,~k=3,~ \gamma=100$ (right plot).
} 
\label{figv23}
\end{center}
\end{figure}

Before we conclude this example, let us show some further solutions in figure \ref{figv23} that fulfill boundary 
conditions on the whole real axis (left plot) and the positive half-axis (right plot).

We observe that these solutions correspond to excited states of the Schr\"odinger equation (\ref{eq3}), which are not 
accessible through the method introduced in \cite{rosu1,rosu2}, but only after its generalization presented in 
section 2.

\section{Concluding remarks} \label{Section remarks}

We have shown the close relationship existing between the zero-mode supersymmetry scheme and the confluent SUSY algorithm of second order. With this relationship we were able to take advantage of some features of the confluent algorithm in order to generalize the results of H. Rosu et. al. (2014) \cite{rosu1, rosu2, rosu3}. One of these features is the fact that there is no need to use only the zero-mode solution in order to perform the transformation. Moreover, we applied the second-order SUSY confluent algorithm to three different equations. First, we took a one dimensional Dirac equation for an inverted harmonic oscillator potential and performed a confluent SUSY transformation in order to generate a new Dirac equation. As a second example a Schr\"odinger equation with constant potential was taken as an initial potential to generate new ones using no zero-mode transformation functions. Finally, we showed how to apply the second order confluent algorithm to the Fokker-Planck equation using not only the ground state but also excited states as transformation functions to generate new Fokker-Planck equations and its solutions.    

\section*{Acknowledgments}
ACA acknowledges CONACYT fellowship 207577.

\end{document}